\documentclass[a4paper]{jpconf}
\def\be{\begin{equation}}
\def\ee{\end{equation}}
\def\bea{\begin{eqnarray}}
\def\eea{\end{eqnarray}}

\usepackage{amsmath}
\usepackage{amsfonts}
\usepackage{amssymb}
\usepackage{graphicx}

\usepackage{graphicx}
\begin{document}
\title{Extended Theories of Gravity with Generalized Energy Conditions}

\author{J~P~Mimoso$^{\ast 1,2}$, F~S~N~Lobo$^{1,2}$ and S~Capozziello$^3$ \\~}

\address{$^{1}$ Departamento de F\'{\i}sica, Faculdade de Ci\^encias da Universidade de Lisboa \\ 
Faculdade de Ci\^encias da Universidade de Lisboa,
Edif\'{i}cio C8, Campo Grande, P-1749-016 Lisbon, Portugal}
\address{$^{2}$ Instituto de Astrof\'{\i}'sica e Ci\^encias do Espa\c co, Faculdade de Ci\^encias da Universidade de Lisboa, 
Edif\'{i}cio C8, Campo Grande, P-1749-016 Lisbon, Portugal}
\address{$^3$ Dipartimento di Fisica, Universit\`{a} di Napoli ``Federico II'', Napoli, Italy
INFN Sez. di Napoli, Compl. Univ. di Monte S. Angelo, Edificio G, Via Cinthia, I-80126, Napoli, Italy.}

\ead{$^1$jpmimoso@fc.ul.pt, $^2$fslobo@fc.ul.pt, $^3$capozzie@na.infn.it}

\begin{abstract}
We address the problem of the energy conditions in modified gravity taking into account the additional degrees of freedom related to scalar fields and curvature invariants. The latter are usually interpreted  as generalized {\it geometrical fluids} that differ in meaning with respect to the  matter fluids generally considered as sources of the field equations. In extended gravity theories the curvature terms are encapsulated in a tensor $H^{ab}$ and a coupling $g(\Psi^i)$ that can be recast as  effective Einstein field equations, with corrections to the energy-momentum tensor of matter. The formal validity of standard energy inequalities does not assure basic requirements such as the attractive nature of gravity, so we argue that the energy conditions have to be considered in a wider sense.
\end{abstract}


\section{Introduction}
Cosmological observations  lead to the introduction of additional ad-hoc concepts like Dark Energy and Dark Matter within
the standard  Einstein theory. However the evasive nature of  these components  may be interpreted as a possible signal of a breakdown of GR on large, infrared  scales \cite{Capo_1}. In such a way, modifications and extensions of GR become a natural alternative if such ``dark'' elements are not found out. Adopting  this viewpoint,  several recent works focussed on the cosmological implications of alternative gravity \cite{Old ETGs,ETGs II,report} since such models may lead to the explanation of the acceleration effect observed in  cosmology \cite{Capozziello:2002,ETGs III} and to the explanation of the missing matter puzzle observed at astrophysical scales \cite{ETGsIV}.

When one considers  modified theories of gravity, at least for a large class of interesting cases, the generalized field equations can be cast in the following form
\begin{equation}
g(\Psi^i)\, \left( G_{ab} +H_ {ab}\right)= 8\pi G\, T_{ab} \; , \label{ETG_EFE1}
\end{equation}
where  $H_{ab}$ is an additional geometrical term with regard to GR that encapsulates the geometrical modifications introduced by the modified theory, and $g(\Psi^i)$ is a factor that modifies the coupling with the matter fields in $T^{ab}$, where $\Psi^i$ generically represents either curvature invariants or other gravitational fields contributing to the dynamics.  In the latter expression, as usual, $G_{ab}\equiv R_{ab}-\frac{1}{2}\,g_{ab}R$ is the  Einstein tensor, defined from the Ricci tensor   $R_{ab}$ and its  trace $R={R^a}_a$,  the curvature scalar.  The field equations govern the interplay between the geometry of the spacetime and the matter content. 
General Relativity (GR) is recovered for $H_{ab}=0$ and $g(\Psi^i)=1$.  

In the Einstein field equations, $G_{ab} = 8\pi G\, T_{ab}$,   there is a clear separation between the left-hand side  that corresponds to the geometry, and the right-hand side where one finds the energy-matter distribution. The underlying idea is that the matter-energy distribution determine  the spacetime curvature, and hence how gravity acts, and it follows that any conditions that we impose on $T_{ab}$ immediately translate into corresponding conditions on the Einstein tensor $G_{ab}$ \cite{Classics}. In this sense, the causal and geodesic  structures of space-time are determined by the matter-energy distribution. In this context, the energy conditions guarantee that the causality principle is respected and suitable physical {\it sources} have to be considered \cite{Classics}. However, in the extended theories (\ref{ETG_EFE1}) not only does the additional term $H_{ab}$ modify the relation between the geometry of gravity and matter, but also the dynamical coupling conveyed by $g(\Psi^i)$ interferes in that interplay.

In this work, we address the problem of the energy conditions in modified gravity \cite{Capozziello:2013vna}. This issue is extremely delicate since a standard approach is to consider the gravitational field equations as  effective Einstein equations. More specifically, the further degrees of freedom carried by these theories \cite{ETGs 5} can be recast as generalized {\it geometrical fluids} that have different meanings with respect to the standard matter fluids generally adopted as sources of the field equations \cite{report}. While standard fluids generally obey standard equations of states, these ``fictitious'' fluids can be related to scalar fields or further gravitational degrees of freedom, as in $f(R)$ gravity. In these cases, the physical properties may be ill-defined and the consequences can be dramatic, since the causal and geodesic structures of the theory could present  serious shortcomings as well as the energy-momentum tensor could not be consistent with the Bianchi identities and the conservation laws.
Thus, we add a cautionary note of the results obtained in the literature \cite{ETG_ec}.
In this work, we adopt the $(-+++)$ signature and $c=1$.

\section{Energy conditions in modified theories of gravity.}
The definition of the energy conditions entails an arbitrary flow which represents a generic observer or a reference frame. In general, we consider a congruence of timelike curves whose tangent 4-vector $W^a$ represents the velocity vector of a family of observers.
 Thus, the energy conditions emerge directly from the geodesic structure of the space-time. More specifically, consider the {\it Raychaudhuri equation}, given by \cite{falco}
\begin{equation}
\dot \theta + \frac{\theta^2}{3}+2\,(\sigma^2-\omega^2) - \dot{W}^a{}_{;a}=- R_ {ab} \, W^a W^b  \;. \label{Ray_1}
\end{equation}
where $2\sigma^2$ is the square of the {\it shear tensor}, $\theta$ is the {\it expansion scalar} and $\omega^2$ is the square of the vorticity tensor.
This Eq. (\ref{Ray_1})  carries only  a geometrical meaning, as the quantities in it are  directly derived from the Ricci identities. It is only when we choose a particular theory that we establish a relation between $R_ {ab} \, W^a W^b $ in Eq. (\ref{Ray_1}), and the energy-momentum tensor describing matter fields \cite{Classics}. One may also consider a null congruence $k^a$ and a vanishing  vorticity $\omega_{ab}=0$, which means that, in GR, it is possible to associate  the null energy condition with the focusing (attracting) characteristic of the spacetime geometry.

Taking into account the diffeomorphism invariance of the matter action,  the covariant conservation of the 
energy-momentum tensor, $\nabla_a T^{ab}=0$, is obtained. Thus, from the contracted Bianchi identities, we derive the following conservation law
\begin{equation}
\nabla_b H^{ab} = - \frac{8\pi G}{g^2} T^{ab}\, \nabla_b g\; .
\end{equation}
The fact that $H^{ab}$ is a geometrical quantity, in the sense that it can be given by geometrical invariants or scalar fields different from ordinary matter fields, implies that the imposition of a specific energy condition on $T^{ab}$ carries an implication for the  combination of $G_ {ab}$ with $H_ {ab}$ and not just for the Einstein tensor.  So we cannot obtain a simple geometrical implication from it, as in GR. For instance, if we assume that the strong energy condition, $T_{ab}\, W^a W^b  \ge \frac{1}{2}T\,W^aW_a$ holds, in GR it would mean, on the one hand, that $R_ {ab} \, W^a W^b  \ge 0$ and, on the other hand, given Eq. (\ref{Ray_1}), that the geodesics are focusing, and  gravity possesses an attractive character. This is one of the assumptions of the singularity theorems of Hawking and Penrose \cite{Classics}. However in the modified gravity context under consideration,  this condition just states that 
\begin{equation}
g(\Psi^i)\,(R_{ab}+H_ {ab}-\frac{1}{2}g_{ab} H)\,W^aW_a \ge 0\; , \label{en_cond_strong_2}
\end{equation}
which does not necessarily imply $R_ {ab} \, W^a W^b  \ge 0$ and hence we cannot straightforwardly conclude that the satisfaction of the strong energy condition (SEC) is synonymous of the attractive nature of gravity in the particular modified theory of gravity under consideration.

In the literature, the term $H^{ab}$ is usually treated as a correction to the energy-momentum tensor, so that the meaning which  is attributed to the energy conditions is the satisfaction of a specific inequality using the combined quantity $T^{ab}_{\rm eff}=T^{ab}/g-H^{ab}$.  It is thus misleading to associate this effective 
energy-momentum tensor to the energy conditions, since they do not emerge only from $T^{ab}$ but from the geometrical quantity $H^{ab}$, which is considered as an  additional energy-momentum tensor. However, if the modified theory of gravity under consideration allows an equivalent description  upon an appropriate conformal transformation, it then becomes justified to associate the transformed $H^{ab}$ to the redefined $T^{ab}$ in the conformally transformed  Einstein frame. 
Several   generalized  theories of gravity  can be redefined as GR plus a number of appropriate fields coupled to matter by means of a conformal transformation in the so-called Einstein frame. This is, for instance, the case for scalar-tensor gravity theories, for $f(R)$ gravity, etc \cite{report}. 

In the original Jordan frame one has a separation between  geometrical terms and  standard matter terms that can be cast as in (\ref{ETG_EFE1}),  where $H_{ab}$  involves a mixture of both the scalar and tensor gravitational fields. 
Upon a suitable conformal transformation we are able to cast the field equations as
$\tilde{G}_ {ab}= 8\pi G\, \tilde{T}_{ab}^{\rm eff}$, 
where $\tilde{T}_{ab}^{\rm eff}=\tilde{T}^M_{ab}+\tilde{T}^\varphi_{ab}$, and it thus makes sense to consider $\tilde{T}_{ab}^{\rm eff}$ as an effective energy-momentum tensor, where $\tilde{T}^M_{ab}$ is the transformed energy-momentum of matter, and $\tilde{T}^\varphi_{ab}$ is an energy-momentum tensor for the redefined scalar field $\varphi$ which is coupled to matter.  
Then 
conclusions about the properties of $\tilde{G}_ {ab}$ such as whether it focuses geodesics directly from those conditions holding on $\tilde{T}_{ab}^{\rm eff}$, ignore the fact that $H_{ab}$ originally possesses a geometrical character, and  may be too hasty if not supported by the physical analysis of the sources. 

If we assume that in this frame the effective energy-momentum tensor $ \tilde{T}^{\rm eff}_{ab}$ satisfies some energy condition, for instance, the null energy condition (NEC), this implies that ${\tilde G}_{ab}$ has to satisfy such a condition. Thus, it is possible to write the Raychaudhuri equation as
\begin{equation}
\frac{{\rm d} \tilde\theta}{{\rm d}v} = - \left[\frac{\tilde\theta^2}{3}+2\tilde\sigma^2 +\tilde R_{ab}\tilde k^a\tilde k^b\right]\; , \label{eq_null}
\end{equation}
which enables us to conclude on the attractive/repulsive character of the given theory of gravity in the Einstein frame. Reversing the conformal transformation,  we can assess, in principle,  what happens in the original frame, namely, the Jordan frame. This  operation  requires to  know   how the kinematical quantities, present in 
Eq. (\ref{eq_null}), transform under a conformal transformation.
This means that if $g_{ab}\to \tilde{g}_{ab}=\Omega^2\,g_{ab}$ and $W^a \to \tilde W^a =\Omega^{-1} W^a$, we have
$\tilde \nabla_a \tilde W_b = \Omega \, \nabla_a W_b +\Omega \, {\gamma^c}_{ab} W_c + W_b\,\nabla_a\Omega $, 
where 
${\gamma^c}_{ab} = \delta^c_a \partial_b\Omega/\Omega+\delta^c_b \partial_a\Omega/\Omega-g_{ab} \partial^c\Omega/\Omega $.

From this result, it follows that we can pass from the Einstein to the Jordan frame by the following transformations
$\tilde{\theta}_{ab} = \Omega\, ( {\theta}_{ab} - \dot\Omega\, h_{ab} )$,
$\tilde{\sigma}_{ab} =  \Omega\,\sigma_{ab} $,
$\tilde\omega_{ab} = \Omega\,\omega_{ab} $,
$\tilde{\theta} = \Omega^{-1}\,( \theta-3\dot\Omega )$,
respectively.
Thus, Eq. (\ref{eq_null}) can finally be written as
$\frac{{\rm d} \tilde\theta}{{\rm d}v}= \frac{\dot\theta}{\Omega^2}-\frac{\theta}{\Omega^2}\frac{\dot\Omega}{\Omega} -\frac{3}{\Omega}\left(\ln \Omega\right)^{..}$.
%
The latter result shows that whereas, in the Einstein frame,  the NEC implies the attractive nature of gravity, a similar implication does not necessarily follow in  the Jordan frame. In fact, $d\tilde\theta/dv\le 0$ only implies that
$\dot\theta \le \frac{\dot\Omega}{\Omega} \theta+{3\Omega}\left(\ln \Omega\right)^{..} $,
and thus it depends on the sign of the term on the right-hand side of the inequality. On the other hand, we see that $\tilde R_{ab}\tilde k^a\tilde k^b\ge 0$ does not necessarily entail
$ R_{ab}k^ak^b \ge 0$. What we do indeed obtain is
\begin{equation}
\left( \Omega^{-2} R_{ab}+   2\nabla_a\nabla_b\ln \Omega +2\nabla_a\ln\Omega\,\nabla_b\ln\Omega \right)\,k^a k^b  \ge 0 \,.
\end{equation}
This emphasizes that if, in one of the conformally related frames, we have  attractive gravity (due to the NEC), in the other frame neither the NEC is simultaneously satisfied, nor, in case it is, this  means that gravity will be straightforwardly attractive. This fact could be extremely relevant in view of identifying a physical meaning of conformal transformations \cite{report,basilakos,magnano}.  


\section{Example of a modified theory of gravity: Scalar-tensor gravity.}

Consider scalar-tensor gravity~\cite{ST} given by the action
\begin{equation}
S=\frac{1}{16\pi} \int \sqrt{-g} d^4x\, \left[\phi R - \frac{\omega(\phi)}{\phi}
\phi_{,a} \phi^{,a} + 2 \phi \lambda(\phi)\right] + S_M \,,
\end{equation}
where $S_M$ is the standard matter part, the gravitational coupling is assumed variable and a self-interaction potential is present. 
Varying this action with respect to the metric $g_{ab}$  and the scalar field $\phi$ yields the field equations (\ref{ETG_EFE1}), with $H_{ab}$ given by
\begin{eqnarray}
H_{ab}&=&-\frac{\omega(\phi)}{\phi^2}\;
\left[\phi_{;a}\phi_{;b} - \frac{1}{2} \, g_{ab}\, 
\phi_{;c}\phi^{;c}\right]
  - \frac{1}{\phi} \left[\phi_{;ab}-g_{ab}
{\phi_{;c}}^{;c}\right]-\lambda(\phi)g_{ab} \,, \label{H-ab_ST}
\end{eqnarray}
and $g(\Psi^i)=\phi$, which we shall assume positive, and
\begin{eqnarray}
&&\Box{\phi}+\frac{{2\phi^2\lambda'(\phi)-2\phi\lambda(\phi)}}
{{2\omega(\phi)+3}}
  = \frac{1}{2\omega(\phi)+3}\;\left[ 8\pi G\, T-\omega'(\phi) 
\phi_{;c}\phi^{;c}
\right] \,,
\end{eqnarray}
where $T\equiv T^{c}{}_{c}$ is the trace of the matter energy-momentum tensor and $G \equiv\frac{2\omega+4}{2\omega+3}\,$ is the gravitational constant normalized to the Newton value. One also requires  the conservation of the matter content $\nabla^a T_{ab}=0$, to preserve  the  equivalence principle. The archetype Brans-Dicke theory is characterized by the restriction of $\omega(\phi)$ being a constant, and of $\lambda=\lambda'=0$. 

The above considerations on the energy conditions apply straightforwardly. In particular, 
Eq. (\ref{en_cond_strong_2}) is easily recovered  like the other energy conditions. 
Since we assume $\phi>0$, we see that the condition $R_{ab}\,W^a\,W^b\ge 0 $ 
becomes 
\begin{eqnarray}
( T_{ab}  - \frac{1}{2}\,g_{ab}\,T)\,W^aW^b \ge \phi\,( H_{ab}  - \frac{1}{2}\,g_{ab}\,H)\,W^aW^b . \label{en_cond_strong_ST1}
\end{eqnarray}
and allows for the focusing of the time-like paths even when a mild  violation of the energy condition occurs. 
Matter may exhibit unusual thermodynamical features, e.g. including negative pressures, and yet gravity remains attractive. Alternatively, 
repulsive gravity may occur for common matter, i.e., for matter that satisfies all the energy conditions. 

In close analogy with the decomposition of the energy-momentum tensor with respect to the vector field $W^a$ \cite{Classics}, one may consider the following 
the decomposition 
of the tensor $H_{ab}$ into the parallel and orthogonal components to the time-like vector flow $W^a$ is given by 
\begin{eqnarray}
H^{ab} &=& H_{||} W^aW^b + H_{\bot} h^{ab} + 2\,H_{\bot}^{(a}\, W^{b)} + H_{\bot}^{<ab>}
  \nonumber   \\
&=& \frac{1}{\phi}\,\left[  \tilde\rho W^aW^b + \tilde p h^{ab} + 2\,\tilde q^{(a}\, W^{b)} + \tilde\pi^{ab}\right] \,.
\end{eqnarray} 
where $H_{||}= H_{ab}\,W^aW^b$, $3H_{\bot} =H_{ab}\,h^{ab}$, $H_{\bot}^{<ab>}= \left(h^{ac}h^{bd}-\frac{1}{3}h^{ab}h^{cd}\right)\,H_{cd}$, and $H_{\bot}^{a}= W^c \,g\,H_{cd}\,h^{ad} $, 
where $H_{||}$ and $H_{\bot}$ are scalars, $H_{\bot}^{a} $ is a vector and $H_{\bot}^{<ab>}$ is a projected trace-free symmetric tensor.
Thus, the inequality (\ref{en_cond_strong_ST1}) may be written as
$(\rho+3p)/\phi-(H_{||}+3H_{\bot}) \ge 0$,
where we have  used the definitions
\begin{eqnarray}
H_{||} &=& - \frac{\omega(\phi)}{2\phi^2}\,\left(3\dot\phi^2-h^{cd}\,\nabla_c\phi\, \nabla_c\phi\right) 
-\frac{1}{\phi}\,h^{cd}\nabla_c\nabla_d\phi +\lambda(\phi) \,, \\
H_{\bot} &=&- \frac{\omega(\phi)}{3\phi^2}\,\left( \frac{\dot\phi^2}{2}-\frac{1}{2}h^{cd}\,\nabla_c\phi\, \nabla_c\phi\right)  
-  \frac{1}{2\phi}\,\left(W^aW^b \, \nabla_c\nabla_d\phi -\frac{1}{3}\,h^{cd}\nabla_c\nabla_d\phi\right) - \lambda(\phi). 
\end{eqnarray}
Thus, $\omega(\phi)$ and $\lambda(\phi)$ define whether gravity is attractive or repulsive in the scalar-tensor cosmological models.
On the other hand, performing 
$g_{ab}\to \bar g_{ab} = (\phi/\phi_\ast)\, g_{ab}$, the  condition for gravity to be attractive  with the redefined Ricci tensor becomes
\begin{equation}
\tilde R_{ab} W^a W^b= \frac{4\pi}{\phi_\ast} \, (\bar\rho+3\bar p) + \frac{8\pi}{\phi_\ast}\, \left[\dot{\varphi}^2-\tilde{V}(\varphi)\right] \ge 0 \,,
\end{equation}
where $\varphi= \int \sqrt{(2\omega+3)/2}\; {\rm d}\ln\phi$ is the redefined scalar field,  $V(\varphi)= \lambda(\phi(\varphi))/\phi(\varphi)$ is the rescaled potential, and $\bar\rho = \rho/\phi^2$, $\bar{p}=p/\phi^2$. 
The role of the  functions $\omega(\phi)$ and $\lambda(\phi)$ underlies the result because the definitions of $\varphi$ and $V(\varphi)$ depend on them. In addition, in the Einstein frame, the matter and the scalar field are interacting with each other as revealed by the scalar field equation
\begin{equation}
\ddot\varphi+\bar{\theta}\dot\varphi= -\frac{\partial V(\varphi)}{\partial \varphi}- \frac{\partial \bar\rho(\varphi,\bar{a})}{\partial \varphi}\; .
\end{equation}
Thus, the dependence of the self-interacting potential $V(\varphi)$, and the coupling $\partial_\varphi \bar{\rho}\propto \alpha(\varphi) a^{-3\gamma}$ is important, where $\alpha=(\sqrt{2\omega+3})^{-1}$, when considering a  perfect fluid with $\bar p=(\gamma-1)\bar\rho$. In a cosmological setting, the interplay of the intervening components such that those which violate the SEC dominate imply that  gravity  exhibits a transition from being attractive into becoming repulsive. This feature is relevant in view of dark energy.
%


\section{Discussion and conclusions.}
We have discussed the formulation and the meaning of the energy conditions in the context of modified theories of gravity. The procedure consists  in  disentangling the further degrees of freedom that emerges with respect to GR and in grouping them as an effective energy-momentum tensor of the form $T^{ab}/g-H^{ab}$ where $g(\Psi^i)$ is the effective coupling and $H^{ab}$ the contribution due to scalar fields and/or curvature invariants of the given modified theory of gravity. 
Formally, the weak, null, dominant and strong energy conditions can be rewritten as in GR. Despite of this analogy, their meaning can be totally  different with respect to GR since the causal structure, geodesic structure and gravitational interaction may be altered.
  
A main role in this analysis is played by recasting the theory  in the Einstein frame.
However, the energy conditions can assume a completely different meaning going back to the Jordan frame and then they could play a crucial role in identifying the physical frame as firstly pointed out in \cite{magnano}. 
On the other hand, geometrical implications change in the two frames since optical scalars like $\sigma$, $\theta$ and $\omega$ can give rise to the convergence or divergence of geodesics. This means that the physical meaning of a given extended theory strictly depends on the energy conditions and initial conditions (in relation to the choice of the source \cite{vignolo}). From an observational point of view, this fact could constitute a formidable tool to test the dark components since deviations from standard GR could be put in evidence.


\ack
SC acknowledges the INFN ({\it iniziative specifiche} TEONGRAV and QGSKY). FSNL is supported by a Funda\c{c}\~{a}o para a Ci\^{e}ncia e Tecnologia Investigador FCT Research contract, with reference IF/00859/2012, funded by FCT/MCTES (Portugal). FSNL and JPM thank   FCT's 
 grants  CERN/FP/123618/201, EXPL/FIS- AST/1608/2013, and PEst-OE/FIS/UI2751/2014.

\section*{References}


\end{document}